\documentclass[eqsecnum,aps,preprintnumbers,groupedaddress,prd,nofootinbib]{revtex4}

\usepackage{amsmath}
\usepackage{enumerate}
\usepackage{amsfonts}
\usepackage{epsfig}

\newcommand{\be}{\begin{equation}}
\newcommand{\ee}{\end{equation}}
\newcommand{\bea}{\begin{eqnarray}}
\newcommand{\eea}{\end{eqnarray}}

\newcommand{\tR}{\tilde{R}}

 \newcommand{\bln}{\begin{align}}
\newcommand{\eln}{\end{align}}
\newcommand{\bst}{\begin{split}}
\newcommand{\est}{\end{split}}
\newcommand{\bi}{\begin{itemize}}
\newcommand{\ei}{\end{itemize}}
\newcommand{\bn}{\begin{enumerate}}
\newcommand{\en}{\end{enumerate}}

\def\AA{{\mathcal A}}
\newcommand{\BB}{{\mathcal B}}

\def\lad{L} %rob {l_{AdS}} changed notation because this is not the ads scale
\def\lg{\lam_{GB}}
\def\ns{N_{\sharp}}
\def\tf{\tilde{f}}
\def\w{\tilde{\om}}
\def\q{\tilde{q}}

\def\ov{\over}
\def\le{\left}
\def\ri{\right}
\def\ha{{1\over 2}}
\def\lam{{\lambda}}
\def\Lam{{\Lambda}}
\def\al{{\alpha}}

\def\vev#1{\langle#1\rangle}

\def\NN{{\cal N}}

\def \lam {\lambda}
\def \om {\omega}
\def \ra {\rightarrow}

\def\ep{{\epsilon}}
\def\apr{{\alpha'}}
\newcommand{\p}{\partial}
\def\LL{{\cal L}}

\def\lam{{\lambda}}

\def\eeq{\end{equation}}

\def\tg{{\tilde g}}
\def\tR{{\tilde R}}
\def\tnab{{\tilde \nabla}}
\def\Rb{{[R^B]}}
\def\htt{{\hat t}}
\def\hx{{\hat x}}
\def\hy{{\hat y}}
\def\hz{{\hat z}}
\def\hr{{\hat r}}

\newcommand{\reef}[1]{(\ref{#1})}
\newcommand{\ssc}{\scriptscriptstyle}

\newcommand{\X}[5]{X^{\ssc (#1)}_{\ #2 #3}{}^{#4 #5}}

\begin{document}

\title {Viscosity Bound Violation in Higher Derivative Gravity}

\preprint{CAS-KITPC/ITP-025}

\preprint{MIT-CTP-3918}

\preprint{SU-ITP-07/22}

\author{Mauro Brigante}
\affiliation{Center for Theoretical Physics,
 Massachusetts
Institute of Technology, \\
Cambridge, MA 02139\\
{\tt brigante@mit.edu}}

\author{Hong Liu}
\affiliation{Center for Theoretical Physics, Massachusetts
Institute of Technology \\
Cambridge, MA 02139\\
{\tt hong\_liu@mit.edu}}

\author{Robert C. Myers}
\affiliation{Perimeter Institute for Theoretical Physics, Waterloo, Ontario N2L 2Y5, Canada\\
and\\
Department of Physics and Astronomy, University of Waterloo, Waterloo, Ontario
N2L 3G1, Canada\\
{\tt rmyers@perimeterinstitute.ca}}

\author{Stephen Shenker}
\affiliation{Department of Physics, Stanford University, Stanford, CA 94305, USA\\
{\tt sshenker@stanford.edu}}

\author{Sho Yaida}
\affiliation{Department of Physics, Stanford University, Stanford, CA 94305, USA\\
{\tt yaida@stanford.edu}}

\date{December, 2007}

\begin{abstract}

Motivated by the vast string landscape, we consider the shear
viscosity to entropy density ratio in conformal field theories
dual to Einstein gravity with curvature square corrections. After
field redefinitions these theories reduce to Gauss-Bonnet gravity,
which has special properties that allow us to compute the shear
viscosity nonperturbatively in the Gauss-Bonnet coupling. By
tuning of the coupling, the value of the shear viscosity to
entropy density ratio can be adjusted to any positive value from
infinity down to zero, thus violating the conjectured viscosity
bound. At linear order in the coupling, we also check consistency
of four different methods to calculate the shear viscosity, and we
find that all of them agree. We search for possible pathologies
associated with this class of theories violating the viscosity
bound.

\end{abstract}

\maketitle
\newpage
\section{Introduction}
\label{intro}

\

The AdS/conformal field theory (CFT) correspondence~\cite{MaldacenaOriginal,GKP,Witten,WittenThermal}
has yielded many important insights into the dynamics of strongly
coupled gauge theories. Among numerous results obtained so far,
one of the most striking is the universality of the ratio of the
shear viscosity $\eta$ to the entropy density
$s$~\cite{Policastro:2001yc,Kovtun:2003wp,Buchel:2003tz,KSSbound}
\begin{equation}
\label{bound} \frac{\eta}{s} = \frac{1}{4\pi}
\end{equation}
for all gauge theories with an Einstein gravity dual in the limit
$N \to \infty$ and $\lam \to \infty$. Here, $N$ is the number of
colors and $\lam$ is the 't Hooft coupling. It was further
conjectured in~\cite{KSSbound} that~(\ref{bound}) is a universal
lower bound [the Kovtun-Starinets-Son (KSS) bound]  for all materials. So far, all known
substances including water and liquid helium satisfy the bound.
The systems coming closest to the bound include the quark-gluon
plasma created at Relativistic Heavy Ion Collider (RHIC)~\cite{Teaney:2003kp,rr1,songHeinz,r2,Dusling:2007gi, Adare:2006nq}
and certain cold atomic gases in the unitarity limit (see
e.g.~\cite{Schafer:2007ib}). $\eta/s$ for pure gluon QCD
slightly above the deconfinement temperature has also been
calculated on the lattice recently~\cite{Meyer:2007ic} and is
about $30 \%$ larger than~(\ref{bound}). See
also~\cite{sakai}. See~\cite{Cohen:2007qr,Cherman:2007fj,Chen:2007jq,Son:2007xw,Fouxon:2007pz}
for other discussions of the bound.

Now, as stated above, the ratio~(\ref{bound}) was obtained for a class of gauge theories whose holographic duals are dictated by classical Einstein gravity (coupled to matter). More generally, string theory (or any
quantum theory of gravity) contains higher derivative corrections from stringy or quantum effects,
inclusion of which will modify the ratio. In terms of gauge theories,
such modifications correspond to $1/\lam$ or $1/N$ corrections. As a concrete example, let us take $\NN=4$ super-Yang-Mills theory, whose dual corresponds to type IIB
string theory on $AdS_5 \times S^5$. The leading order correction in $1/\lam$ arises from stringy corrections to the low-energy effective action of type IIB supergravity, schematically of the form $\apr^3 R^4$.
The correction to $\eta/s$ due to such a term was calculated
in~\cite{BLS,Benincasa:2005qc}. It was found that the correction is positive, consistent with the
conjectured bound.

In this paper, instead of limiting ourselves to
specific known string theory corrections, we explore the modification of $\eta/s$ due to
generic higher derivative terms in the holographic gravity dual. The reason is partly pragmatic: other
than in a few maximally supersymmetric circumstances, very little
is known about forms of higher derivative corrections generated in string theory. Given the vastness of the string
landscape~\cite{landscape}, one expects that generic corrections do
occur. Restricting to the gravity sector in $AdS_5$, the leading order
higher derivative corrections can be written as\footnote{Our
conventions are those of \cite{Carroll}. In this section we suppress Gibbons-Hawking surface terms.}
 \be \label{epr}
I= {1 \ov 16 \pi G_N} \int d^5 x \, \sqrt{- g} \le(R - 2 \Lambda +
\lad^2  \le(\al_1  R^2+  \al_2 R_{\mu \nu} R^{\mu \nu}+\al_3 R^{\mu
\nu\rho \sigma} R_{\mu \nu \rho\sigma} \ri)\ri) \ ,
 \ee
where $\Lam = -{6 \ov \lad^2}$ and for now we assume that $\al_i \sim
{\apr \ov \lad^2} \ll 1$. Other terms with additional derivatives
or factors of $R$ are naturally suppressed by higher powers of ${\apr \ov
\lad^2}$. String loop (quantum) corrections can also generate such terms,
but they are suppressed by powers of $g_s$ and we will consistently neglect them by taking $g_s \rightarrow 0$ limit.\footnote{Note that to calculate $g_s$ corrections, all the light fields must be taken into account. In addition, the calculation of $\eta/s$ could be more subtle once we begin to include quantum effects.} To lowest order
in $\al_i$ the correction to $\eta/s$ will be a linear
combination of $\al_i$'s, and the viscosity bound
is then violated for one side of the half-plane.
Specifically, we will find
 \be
 {\eta \ov s} = {1 \ov 4 \pi} \le(1 - 8 \al_3 \ri) + O(\al_i^2)
 \ee
and hence the bound is violated for $\al_3>0$. Note that the above
expression is independent of $\al_1$ and $\al_2$. This can be
inferred from a field redefinition argument (see
Sec.\ref{ap:fR}).

How do we interpret these violations? Possible scenarios are:

\begin{enumerate}

\item The bound can be violated. For example, this scenario would be realized if one explicitly finds a well-defined string theory on $AdS_5$ which generates a stringy correction with $\al_3>0$. (See~\cite{new} for a plausible counterexample to the KSS bound.)

\item The bound is correct (for example, if one can prove it using a field
theoretical method), and a bulk gravity theory with $\al_3>0$ cannot have a well-defined boundary CFT dual.

 \begin{enumerate}

\item The bulk theory is manifestly
inconsistent as an effective theory. For example, it could
violate bulk causality or unitarity.

\item It is impossible to generate such a low-energy effective
classical action from a consistent quantum theory of gravity. In
modern language we say that the theory lies in the swampland of
string theory.

 \end{enumerate}

\end{enumerate}

Any of these alternatives, if realized, is interesting. Needless
to say, possibility 1 would be interesting. %While there is clear
%evidence that for QCD $\eta/s$ is bounded from above, 
Given that recent analyses from RHIC
data~\cite{rr1,songHeinz,r2,Dusling:2007gi,Adare:2006nq} indicate 
the $\eta/s$ is close to (and could be even smaller than) the bound, 
%are important steps
%toward being able to bound it from below. 
this further motivates
to investigate the universality of the KSS bound in holographic
models.

Possibility 2(a) should help clarify the physical origin of the
bound by correlating bulk pathologies and the violation of the
bound. Possibility 2(b) could provide powerful tools for
constraining possible higher derivative corrections in the string
landscape. Note that while there are some nice no-go theorems
which rule out classes of nongravitational effective field
theories \cite{AADNR} (also see \cite{AKS}), the generalization of
the arguments of~\cite{AADNR} to gravitational theories is subtle
and difficult. Thus, constraints from AdS/CFT based on the
consistency of the boundary theory would be valuable.

In investigating the scenarios above, Gauss-Bonnet (GB) gravity will
provide a useful model. Gauss-Bonnet gravity, defined by the
classical action of the form~\cite{Zwiebach}
\begin{equation}
\label{action} I = \frac{1}{16\pi G_N} \mathop\int{d^{5}x \,
\sqrt{-g} \, \le[R-2\Lambda+ {\lg \ov 2} \lad^2
(R^2-4R_{\mu\nu}R^{\mu\nu}+R_{\mu\nu\rho\sigma}R^{\mu\nu\rho\sigma})
\ri]} \ ,
\end{equation}
has many nice properties that are absent for theories with more general ratios of the $\al_i$'s. For example, expanding around flat Minkowski space, the metric fluctuations
have exactly the same quadratic kinetic terms as those in
Einstein gravity. All higher derivative terms
cancel~\cite{Zwiebach}. Similarly, expanding around the
AdS black brane geometry, which will be the main focus of the
paper, there are also only second derivatives on the
metric fluctuations. Thus small metric fluctuations can be
quantized for finite values of the parameter
$\lg$.\footnote{Generic theories in~(\ref{epr}) contain four
derivatives and a consistent quantization is not possible other
than treating higher derivative terms as perturbations.}
Furthermore, crucial for our investigation is its remarkable feature of solvability: sets of
exact solutions to the classical equation of motion have been
obtained \cite{BD,Cai} and the exact form of the Gibbons-Hawking
surface term is known \cite{Myers}.

Given these nice features of Gauss-Bonnet gravity, we will venture outside the regime of the perturbatively corrected Einstein gravity and study the theory with finite values of $\lg$.
To physically motivate this, one could envision that somewhere in the string landscape $\lg$ is large but all the other higher derivative corrections are small.
One of the main results of the paper is a value of $\eta/s$ for the CFT dual of Gauss-Bonnet gravity, \emph{nonperturbative} in $\lg$:\footnote{We have also computed the value of $\eta/s$ for Gauss-Bonnet gravity
for any spacetime dimension $D$ and the expression is given
in~(\ref{final}).}
\begin{equation} \label{advertise}
\frac{\eta}{s}=\frac{1}{4\pi}[1- 4 \lg].
\end{equation}
We emphasize that this is not just a linearly corrected value.
In particular, the viscosity bound is badly violated as $\lg \rightarrow \frac{1}{4}$.
As we will discuss shortly, $\lg$ is bounded above by $\frac{1}{4}$ for the theory to have a boundary CFT, and $\eta/s$ never decreases beyond $0$.

Given the result~(\ref{advertise}) for Gauss-Bonnet, if the possibility 2(a) were correct, we would expect that pathologies would become easier to discern in the limit where $\eta/s\rightarrow 0$. We will investigate this line of thought in Sec.\ref{gravitoncone}. On the other hand, thinking along the line of possibility 1, the Gauss-Bonnet theory with $\lg$ arbitrarily close to $\frac{1}{4}$ may have a concrete realization in the string landscape. In this case, there exists no lower bound for $\eta/s$, and investigating the CFT dual of Gauss-Bonnet theory should clarify how to evade the heuristic mean free path argument for the existence of the lower bound (presented in, e.g., \cite{KSSbound}).

The plan of the paper is as follows. In Sec.\ref{pre}, we review
various properties of two-point correlation functions and outline
the real-time AdS/CFT calculation of the shear viscosity. We then
explicitly calculate the shear viscosity for Gauss-Bonnet theory
in Sec.\ref{shear}. In Sec.\ref{gravitoncone}, we seek possible
pathologies associated with theories violating the viscosity
bound. There, we will find a curious new metastable state for
large enough $\lg$. Finally in Sec.\ref{discussion}, we conclude
with various remarks and speculations. To make the paper fairly
self-contained, various appendices are added. In particular,
quasinormal mode calculations of the shear viscosity are
presented in Appendix~\ref{ap:so} and one using the membrane
paradigm in Appendix~\ref{junk}.

\section{Shear viscosity in $R^2$ theories: preliminaries} \label{pre}

\subsection{Two-point correlation functions and viscosity} \label{meds}

Let us begin by collecting various properties of two-point
correlation  functions,
following~\cite{Policastro:2002se,Policastro:2002tn,KovtunEV} (see
also~\cite{Son:2007vk}). Consider retarded two-point correlation
functions of the stress energy tensor $T_{\mu \nu}$ of a CFT in
$3+1$-dimensional Minkowski space at a finite temperature $T$
 \be \label{ttC}
 G_{\mu\nu,\al \beta} (\omega,\vec q) = - i\int dt d\vec x e^{ i
\omega t- i \vec q\cdot \vec x} \theta( t) \vev{\le[T_{\mu
\nu}(t,\vec x) , T_{\al \beta} (0,0)\ri]}.
 \ee
They describe linear responses of the system to small disturbances.
It turns out that various components of~(\ref{ttC}) can be expressed in terms of
three independent scalar functions. For example, if we take spatial
momentum to be $\vec q = (0,0 ,q)$, then
 \be
 G_{12,12}= \ha G_3 (\om, q) , \ \ \ \ \ G_{13,13} = \ha{\om^2 \ov \om^2 -q^2}
 G_1 (\om, q), \ \ \ \ \ G_{33,33} = {2 \ov 3} {\om^4 \ov (\om^2 -q^2)^2} G_2 (\om,
 q),
% \qquad G_{00,00} = {2 \ov 3} {q^4 \ov (\om^2 -q^2)^2} G_2 (\om, q)
 \ee
and so on. At $\vec q=0$ all three function $G_{1,2,3} (\om,0)$
are equal to one another as a consequence of rotational
symmetry.

When $\om, |\vec q| \ll T$ one expects the CFT plasma to be described by hydrodynamics.
The scalar functions $G_{1,2,3}$ encode the hydrodynamic behavior of shear, sound, and transverse modes, respectively.
More explicitly, they have the following properties:
   \begin{itemize}
   \item $G_1$ has a simple diffusion pole at $\omega= - i D q^2$, where
    \be \label{shearA}
    D= {\eta \ov \ep + P} = {1 \ov T} { \eta \ov  s}
    \ee
    with $\ep$ and $s$ being the energy and entropy density, and $P$ the pressure of the
    gauge theory plasma.
   \item $G_2$ has a simple pole at $\omega= \pm c_s q - i \Gamma_{s} q^2$, where $c_s$ is
   the speed of sound and $\Gamma_{s}$ is the sound damping
   constant, given by (for conformal theories)
 \be \label{sound}
 \Gamma_s = {2 \ov 3T}{\eta \ov s}
 \ee
   \item $\eta$ can also be obtained from $G_{1,2,3}$ at zero spatial momentum by
the Kubo formula, e.g.,
 \be \label{rrp}
 \eta= \lim_{\omega\ra 0} {1\ov \omega} {\rm Im} G_{12,12} (\om,0) % =
%\lim_{\omega\ra 0}  {1 \ov 2 \om} {\rm Im} G_3 (\om,0)
 \ee

 \end{itemize}
Equations (\ref{shearA})--(\ref{rrp}) provide three independent ways of extracting $\eta/s$.
We provide calculations utilizing the first two in Appendix~\ref{ap:so}.
A calculation utilizing the Kubo formula (\ref{rrp}) is easier, and we will explicitly implement it for Gauss-Bonnet theory in Sec.\ref{shear}.
In the next subsection, we outline how to obtain retarded two-point functions within the framework of the real-time AdS/CFT correspondence.

\subsection{AdS/CFT calculation of shear viscosity: Outline} \label{adscft}

The stress tensor correlators for a boundary CFT described by
(\ref{epr}) or (\ref{action}), can be computed from gravity as
follows. One first finds a black brane solution (i.e. a black hole
whose horizon is ${\bf R}^3$) to the equations of motion of
(\ref{epr}) or (\ref{action}). Such a solution describes the
boundary theory on ${\bf R}^{3,1}$ at a temperature $T$, which can
be identified with the Hawking temperature of the black brane. The
entropy and energy density of the boundary theory are given by the
corresponding quantities of the black brane. The fluctuations of
the boundary theory stress tensor are described in the gravity
language by small metric fluctuations $h_{\mu \nu}$ around the
black brane solution. In particular, after taking into account of
various symmetries and gauge degrees of freedom, the metric
fluctuations can be combined into three independent scalar fields
$\phi_a, a=1,2,3$, which are dual to the three functions $G_a$ of the
boundary theory.

To find $G_a$, one could first work out the bulk two-point
retarded function for $\phi_a$ and then take both points to the
boundary of the black brane geometry. In practice it is often more
convenient to use the prescription proposed in~\cite{SS}, which
can be derived from the real-time AdS/CFT
correspondence~\cite{HS}. Let us briefly review it here:

\bn

\item  Solve the linearized equation of motion for $\phi_a (r; k)$ with the
following boundary conditions:

\bn

\item Impose the infalling boundary condition at the horizon. In other words, modes with timelike momenta should be
falling into the horizon and modes with spacelike momenta should
be regular.

\item Take $r$ to be the radial direction of the black brane geometry
with the boundary at $r=\infty$. Require
 \be \label{bdw}
 \phi_a (r; k)|_{r= {1 \ov \ep}} = J_a (k),   \qquad k = (\om, q),
 \ee
where $\ep \to 0$ imposes an infrared cutoff near the infinity
of the spacetime and $J_a (k)$ is an infinitesimal boundary source for the
bulk field $\phi_a(r; k)$.

\en

\item Plug in the above solution into the action, expanded to quadratic order in $\phi_a (r; k)$.
It will reduce to pure surface contributions.
The prescription instructs us to pick up only the contribution from the boundary at $r={1 \ov \ep}$.
The resulting action can be written as
\begin{equation}\label{Sbd}
S = - \ha \int\!{d^4k\ov (2\pi )^4}\,
  J_a (-k) {\cal F}_a (k,r) J_a (k) \Big|_{r={1 \ov \ep}}\ .
\end{equation}
Finally the retarded function $G_a (k)$ in momentum space for the
boundary field dual to $\phi_a$ is given by
 \be \label{eo}
 G_a (k) = \lim_{\ep \to 0} {\cal F}_a (k,r)\Big|_{r={1 \ov \ep}} \
 .
 \ee

\en
Using the Kubo formula~(\ref{rrp}), we can get the shear viscosity by studying a mode $\phi_3$ with $\vec q=0$ in the low-frequency limit $\om \rightarrow 0$. We will do so in the next section. Alternatively, using (\ref{shearA}) or (\ref{sound}), we can read off the viscosity from pole structures of retarded two-point functions. Such a calculation is a bit more involved and will be performed in Appendix~\ref{ap:so}.

The above prescription for computing retarded functions in AdS/CFT
works well if the bulk scalar field has only two derivatives as in
Gauss-Bonnet case~(\ref{action}). If the bulk action contains more
than two derivatives, complications could arise even if one treats
the higher derivative parts as perturbations. For example, one
needs to add Gibbons-Hawking surface terms to ensure a
well-defined variational problem. A systematic prescription for
doing so is, however, not available at the moment beyond the
linear order. Thus there are potential ambiguities in implementing
(\ref{eo}).\footnote{In~\cite{BLS}, such additional terms do not
appear to affect the calculation at the order under discussion
there.} Clearly these are important questions which should be
explored more systematically. At the $R^2$ level, as we describe
below in Sec.\ref{ap:fR}, all of our calculations can be reduced
to the Gauss-Bonnet case in which these potential complications do
not arise.

\subsection{Field redefinitions in $R^2$ theories} \label{ap:fR}

We now show that to linear order in $\al_i$, $\eta/s$
for~(\ref{epr}) is independent of $\al_1$ and $\al_2$. It is well
known that to linear order in $\al_i$, one can make a field
redefinition to remove the $R^2$ and $R_{\mu \nu}R^{\mu\nu}$ term
in~(\ref{epr}). More explicitly, in~(\ref{epr}) set $\al_3 =0$ and
take
 \be \label{eep}
  g_{\mu \nu} = \tilde g_{\mu \nu} + \al_2 \lad^2 \tilde R_{\mu \nu} - {\lad^2 \ov 3}
  (\al_2
+ 2 \al_1 ) \tilde g_{\mu \nu} \tilde R,
 \ee
where $\tilde R$ denotes the Ricci scalar for $\tilde g_{\mu\nu}$
and so on. Then (\ref{epr}) becomes
 \be \label{newz}
 I  =  {1 \ov 16 \pi G_N} \int \sqrt{- \tilde g} ((1+ {\cal K})
 \tilde R - 2  \Lambda  ) + O(\al^2)
  = {1 +  {\cal K}\ov 16 \pi G_N} \int \sqrt{- g} (   \tilde R - 2 \tilde \Lambda ) +
  O(\al^2)
 \ee
with \be {\cal K}= {2 \Lam \lad^2 \ov 3} \le(5 \al_1 + \al_2 \ri) , \qquad
 \tilde \Lam = {\Lam \ov 1 + {\cal K}} \ .
\ee It follows from (\ref{eep}) that a background solution
$g^{(0)}$ to (\ref{epr}) (with $\al_3=0$) is related to a solution
$\tilde g^{(0)}$ to (\ref{newz}) by
 \be \label{sba}
 ds^2_0  = A^2 \tilde{ds}^2_0, \qquad A = 1- { {\cal K} \ov 3} \ .
 \ee
The scaling in (\ref{sba}) does not change the background Hawking
temperature.  The diffusion pole~(\ref{shearA}) calculated using
~(\ref{newz}) around $\tilde g^{(0)}$ then gives the standard
result $D = {1 \ov 4 \pi T}$~\cite{Policastro:2002se}.
 Thus we conclude that $\eta/s = {1 \ov
4 \pi}$ for (\ref{epr}) with $\al_3=0$. Then to linear order
in $\al_i$,
 $\eta/s$ can only
depend on $\al_3$. To find this dependence,  it is convenient to
work with the Gauss-Bonnet theory~(\ref{action}). Gauss-Bonnet
gravity is not only much simpler than~(\ref{epr}) with generic
$\al_3 \neq 0$, but also contains only second derivative terms in
the equations of motion for $h_{\mu \nu}$, making the extraction
of boundary correlators unambiguous.

\section{Shear Viscosity for Gauss-Bonnet Gravity}
\label{shear}

In this section, after briefly reviewing the thermodynamic properties of the black brane solution, we compute the shear viscosity for Gauss-Bonnet gravity~(\ref{action}) nonperturbatively in $\lg$.
Here, we follow the outline presented in the previous section, with the Kubo formula (\ref{rrp}) in mind.
In Appendix~\ref{ap:so}, we extract $\eta/s$  from the shear channel~(\ref{shearA}) and the sound channel~(\ref{sound}) (perturbatively in $\lg$).
There we also find that the sound velocity remains at the conformal value $c_s^2 = {1 \ov 3}$ as it should.
In Appendix~\ref{junk}, we provide a membrane paradigm calculation, again nonperturbatively in $\lg$.
All four methods give the same result.

\subsection{Black brane geometry and thermodynamics}

Exact solutions and thermodynamic properties of black objects in
Gauss-Bonnet gravity~(\ref{action}) were discussed
in~\cite{Cai} (see also \cite{Nojiri:2001aj, Cho:2002hq,Neupane:2002bf,Neupane:2003vz}). Here we summarize some features relevant for our
discussion below. The black brane solution can be written as
\begin{equation}
\label{bba} ds^2=-f(r)\ns^2dt^2
+\frac{1}{f(r)}dr^2+\frac{r^2}{\lad^2}
 \le(\mathop\sum_{i=1}^{3}dx_i^2 \ri),
\end{equation}
where
\begin{equation}
\label{perturb} f(r)=\frac{r^2}{\lad^2}\frac{1}{2\lg}
\le[1-\sqrt{1-4\lg\le(1-\frac{r_{+}^{\,4}}{r^4}\ri)} \ri] \ .
\end{equation}
In (\ref{bba}), $\ns$ is an arbitrary constant which specifies the
speed of light of the boundary theory. Note that as $r \to \infty$,
\be \label{Ade}
 f(r) \to {r^2 \ov a^2 \lad^2}, \qquad {\rm with} \qquad
  a^2 \equiv %\frac{2\lg}{1-\sqrt{1-4\lg}}
  \ha\le(1+\sqrt{1-4\lg}\ri)\ .
 \ee
It is straightforward to see that the AdS curvature scale of these
geometries is $a \lad$.\footnote{Here we note that the Gauss-Bonnet
theory also admits another background with the curvature scale
$\tilde{a}\,\lad$ where $\tilde{a}^2=\ha\le(1-\sqrt{1-4\lg}\ri)$.
Even though this remains an asymptotically AdS solution for $\lg>0$,
we do not consider it here because this background is unstable and
contains ghosts \cite{BD}.} If we choose $\ns = a$, then the boundary
speed of light is unity. However, we will leave it unspecified in
the following. We assume that $\lg\leq\frac{1}{4}$. Beyond this
point, (\ref{action}) does not admit a vacuum AdS solution, and
cannot have a boundary CFT dual. In passing, we note that while the
curvature singularity occurs at $r=0$ for $\lg \geq 0$, it shifts to
$r =r_+ \le(1-{1 \ov 4 \lg}\ri)^{-\frac{1}{4}}$ for $\lg<0$.

The horizon is located at $r=r_{+}$ and the Hawking temperature,
entropy density, and  energy density of the black brane are
\footnote{Note that for {\it planar} black branes in Gauss-Bonnet
theory, the area law for entropy still holds \cite{oldy}. This is
not the case for more general higher-derivative-corrected black
objects.}
\begin{equation}
\label{temperature} T =\ns \frac{r_{+}}{\pi \lad^2},
\end{equation}
\begin{equation} \label{entr}
s =\frac{1}{4G_{N}}\le(\frac{r_{+}}{\lad}\ri)^3
=\frac{(\pi \lad)^3}{4G_{N}}\frac{(T)^3}{\ns^3},
\qquad {\ep} = {3 \ov 4} T s \ .
\end{equation}
If we fix the boundary theory temperature $T$ and the speed of light to be unity (taking $\ns=a$), the entropy and energy density are monotonically
increasing functions of $\lg$, reaching a maximum at $\lg={1 \ov 4}$
and going to zero as $\lg \to -\infty$.

To make our discussion self-contained, in Appendix~\ref{ap:ther}, we compute the free
energy of the black brane and derive the entropy density. In
particular, we show that the contribution from the Gibbons-Hawking
surface term to the free energy vanishes.

\subsection{Action and equation of motion for the scalar channel} \label{scalar}

To compute the shear viscosity, we now study small metric fluctuations $\phi = h^1_{\ 2}$ around the black brane background of the form
\begin{equation}
ds^2=-f(r)\ns^2dt^2+\frac{1}{f(r)}dr^2+\frac{r^2}{\lad^2}
\le(\mathop\sum_{i=1}^{3}dx_i^2 + 2 \phi(t,\vec x,  r)dx_1 d x_2 \ri) \ .
\end{equation}
We will take $\phi$ to be independent of $x_1$ and $x_2$ and write
\begin{equation}
\phi(t, \vec x, r)=\mathop\int\frac{d\om dq}{(2\pi)^{2}}\, \phi(r;k)
\, e^{-i\omega t + i q x_3}, \quad k = (\om, 0, 0, q) ,\ \ \phi
(r; -k)=\phi^*(r; k) \ .
\end{equation}
For notational convenience, let us introduce
\begin{equation} \label{sDef}
z=\frac{r}{r_{+}},\ \ \tilde{\omega}=\frac{\lad^2}{r_{+}}\omega,\
\quad \tilde{q}=\frac{\lad^2}{r_{+}}q, \qquad
\tilde{f}=\frac{\lad^2}{r_{+}^2}f =  {z^2 \ov 2 \lg} \le(1 -
\sqrt{1-4 \lg + {4 \lg \ov z^4}} \ri).
\end{equation}
Then, at quadratic order, the action for $\phi$ can be written as
 \begin{eqnarray} \label{pee}
 S&=&\int{dk_1 dk_2 \ov (2 \pi)^2}S(k_1, k_2) \ \ \ {\rm with} \cr
 S(k_1=0, k_2=0)&=&-\ha C \int dz {\frac{d\om dq}{(2\pi)^{2}}} \, \le( K (\p_z \phi)^2 - K_2 \phi^2  +
 \p_z (K_3 \phi^2) \ri),
 \end{eqnarray}
where
 \be \label{vds}
 C = {1 \ov 16 \pi G_N} \le(\ns r_+^4\ov \lad^5\ri), \ \ K= z^2 \tf (z - \lg
 \p_z\tf), \ \ \ K_2 = K {\tilde \om^2 \ov \ns^2 \tf^2} - \tilde
 q^2 z \le(1- \lg \p_z^2\tf \ri) \ ,
 \ee
and $\phi^2$ should be understood as a shorthand notation for
$\phi(z;k) \phi (z,-k)$.
Here, $S$ is the sum of the bulk action (\ref{action}) and the associated Gibbons-Hawking surface term \cite{Myers}.
The explicit expression for $K_3$ will not be important for our subsequent discussion.

The equation of motion following from (\ref{pee}) is\footnote{An easy way to get the quadratic action~(\ref{pee}) is to first obtain the linearized equation of motion and then read off $K$ and $K_2$ from it.}
 \be \label{eom}
 K \phi'' + K' \phi' + K_2 \phi =0 \ ,
 \ee
where primes indicate partial derivatives with respect to $z$.
Using the equation of motion, the action~(\ref{pee}) reduces to the surface contributions as advertised in Sec.\ref{adscft},
 \be \label{rrk}
 S(k_1=0, k_2=0) = -\ha C \int {\frac{d\om dq}{(2\pi)^{2}}} \, \le(K \phi' \phi
 + K_3 \phi^2 \ri)|_{{\rm surface}} \ .
 \ee
The prescription described in Sec.\ref{adscft} instructs us to
pick up the contribution from the boundary at
$z\rightarrow\infty$. Here, the term proportional to $K_3$ will
give rise to a real divergent contact term, which is discarded.

A curious thing about (\ref{pee}) is that for all values of $z$,
both $K$ and $K_2$ (but not $K_3$) are proportional to ${1 \ov 4}
- \lg$.\footnote{This can be seen by using the following equation
in $K$ and $K_2$ \be \tf'(z) = {2 z (2z^2 -\tf) \ov z^2 - 2 \lg
\tf} \ . \ee } Thus other than the boundary term the whole action
(\ref{pee}) vanishes identically at $\lg = {1 \ov 4}$.
Nevertheless, the equation of motion (\ref{eom}) remains
nontrivial in the limit $\lg \to {1 \ov 4}$ as the ${1 \ov 4} - \lg$ factor cancels
out. Note that the correlation function does not necessarily go to
zero in this limit since it also depends on
the behavior of the solution to~(\ref{eom}) and the limiting
procedure~(\ref{rrk}). As we will see momentarily, as least in the
small frequency limit it does become zero with a vanishing shear
viscosity.

\subsection{Low-frequency expansion and the viscosity}
\label{solution}

General solutions to the equation of motion~(\ref{eom}) can be written as
\be
\phi(z; k)=a_{in}(k)\phi_{in}(z; k)+a_{out}(k)\phi_{out}(z; k) \ ,
\ee
where $\phi_{in}$ and $\phi_{out}$ satisfy infalling and outgoing boundary conditions at the horizon, respectively.
They are complex conjugates of each other, and we normalize them by requiring them to approach $1$ as $z\to \infty$.
Then, the prescription of Sec.\ref{adscft} corresponds to setting
\be \label{explicitBC}
a_{in}(k)=J(k)\ , \qquad a_{out}(k)=0 \ ,
\ee
where $J(k)$ is an infinitesimal boundary source for the bulk field $\phi$.

More explicitly, as $z \to 1$, various functions in (\ref{eom}) have the following behavior
 \be
{K_2 \ov K} \approx {\w^2 \ov 16 \ns^2 (z-1)^2 } + O((z-1)^{-1})+O(\q^2),
\qquad  {K' \ov K} = {1 \ov z-1} + O(1) \ .
 \ee
It follows that near the horizon $z=1$, equation (\ref{eom}) can
be solved by (for $\vec q =0$)
 \be
\phi (z) \sim (z-1)^{\pm {i \w \ov 4 \ns}} \sim (z-1)^{\pm {i \om
\ov 4 \pi T}}
 \ee
with the infalling boundary condition corresponding to the
negative sign. To solve (\ref{eom}) in the small frequency limit,
it is convenient to write
\begin{equation} \label{anse}
\phi_{in} (z; k)=e^{-i \le({\w \ov 4 \ns}\ri) {\rm ln}\le(\frac{a^2
\tilde{f}}{z^2}\ri)} \le(1-i\frac{\w} {4
\ns}g_1(z)+O(\tilde{\omega}^2, \q^2)\ri),
\end{equation}
where we require $g_1 (z)$ to be nonsingular at the horizon $z=1$.
We show in Appendix~\ref{ap:solo} that $g_1$ is a nonsingular function with the large $z$
expansion
 \be \label{lowEatCFT}
 g_1 (z) = {4 \lg \ov \sqrt{1-4 \lg}} {a^2 \ov z^4} + O(z^{-8}) \
 .
 \ee
Therefore, with our boundary conditions (\ref{explicitBC}), we find
 \be \label{asu}
 \phi (z;k) =J(k)\le[ 1 +  {i \w \ov 4 \ns} a^2 \sqrt{1-4 \lg} \le({1 \ov z^4}
 + O(z^{-8}) \ri) + O(\w^2, \q^2)\ri].
 \ee
This is the right asymptotic behavior for the bulk field $\phi$
describing metric fluctuations since the CFT stress tensor has
conformal dimension 4.

Plugging~(\ref{asu}) into (\ref{rrk}) and using the expressions for
$C$ and $K$ in (\ref{vds}), the prescription described in Sec.\ref{adscft} gives
 \be
 {\rm Im} G_{12,12} (\om,0)=\omega{1 \ov 16 \pi G_N} \le(r_+^3\ov \lad^3\ri)  (1-4
 \lg) +O(\omega^2).
 \ee
Then, the Kubo formula~(\ref{rrp}) yields
 \be \label{ets}
 \eta = {1 \ov 16 \pi G_N} \le(r_+^3\ov \lad^3\ri)  (1-4
 \lg).
 \ee
Finally, taking the ratio of (\ref{ets}) and (\ref{entr}) we find that
 \be \label{ror}
 {\eta \ov s} = {1 \ov 4 \pi} (1-4
 \lg).
 \ee
This is \emph{nonperturbative} in $\lg$. Especially,
the linear correction is the only nonvanishing term.\footnote{It would be interesting to find an explanation for vanishing of higher order corrections.}

We now conclude this section with various remarks:

\begin{enumerate}

\item Based on the field redefinition argument presented in
Sec.\ref{ap:fR}, one finds from (\ref{ror}) that for (\ref{epr}),
 \be \label{oror}
 {\eta \ov s} = {1 \ov 4 \pi} \le(1 - 8 \al_3 \ri) + O(\al_i^2).
 \ee
We have also performed an independent calculation of $\eta/s$
(without using field redefinitions) for~(\ref{epr}) using all
three methods outlined in Sec.\ref{meds} and
confirmed~(\ref{oror}).

\item The ratio $\eta/s$ dips below the viscosity bound for $\lg
> 0$ in Gauss-Bonnet gravity and for $\al_3 > 0$ in~(\ref{epr}).
In particular, the shear viscosity approaches zero as $\lg \to {1 \ov 4}$ for Gauss-Bonnet.
 Note that the whole off-shell action becomes zero in this limit. It is
likely the on-shell action also vanishes, implying that the
correlation function could become identically zero in this limit.

\item Fixing the temperature $T$ and the boundary speed of light
to be unity, as we take $\lg \to -\infty$, $\eta \sim (-\lg)^{1
\ov 4} \to \infty$. In contrast the entropy density decreases as
$s \sim (-\lg)^{-{3 \ov 4}} \to 0$.

\item The shear viscosity of the boundary conformal field theory
is associated with absorption of transverse modes by the black
brane in the bulk. This is a natural picture since the shear
viscosity measures the dissipation rate of those fluctuations: the
quicker the black brane absorbs them, the higher the dissipation
rate will be.
For example, as $\lg \rightarrow-\infty$, $\eta/s$ approaches
infinity; this describes a situation where every bit of the black
brane horizon devours the transverse fluctuations very quickly.
In this limit the curvature singularity at $z = \le(1-{1 \ov 4 \lg}\ri)^{-\frac{1}{4}}$ approaches the horizon and the tidal force near the horizon becomes strong.
On the other hand, as $\lg \to {1 \ov 4}$,
$\eta/s\rightarrow0$ and the black brane very slowly absorbs transverse modes.\footnote{We note that for
$\lg=\frac{1}{4}$ in $4+1$ spacetime dimension, the radial direction of the background
geometry resembles a ${\rm Ba\tilde{n}ados}$-Teitelboim-Zanelli (BTZ) black brane.}

\item The calculation leading to (\ref{ror}) can be generalized to
general $D$ spacetime dimensions and one finds for $D\geq4+1$\footnote{For general
dimensions we use the convention
  \begin{equation}
S = \frac{1}{16\pi G_N} \mathop\int{d^{D}x \, \sqrt{-g} \,
\le[R-2\Lambda+ \alpha_{GB} \lad^2
(R^2-4R_{\mu\nu}R^{\mu\nu}+R_{\mu\nu\rho\sigma}R^{\mu\nu\rho\sigma})
\ri]} \
\end{equation}
with $\Lam = - {(D-1) (D-2) \ov 2 \lad^2}$ and $\lg = (D-3)(D-4)
\alpha_{GB}$.}
 \begin{equation}
\label{final} \frac{\eta}{s}=\frac{1}{4\pi}\le[1-2\frac{(D-1)}{(D-3)}\lg \ri] \ .
\end{equation}
Here again $\lg$ is bounded above by ${1 \ov 4}$. Thus for $D >
4+1$, $\eta$ never approaches zero within Gauss-Bonnet theory. For
$D=3+1$ or $2+1$, in which case the Gauss-Bonnet term is
topological, there is no correction to $\eta/s$.

\item In Appendix~\ref{junk}, we obtain the same result
(\ref{ror}) using the membrane paradigm \cite{Kovtun:2003wp}. Thus
when embedded into the AdS/CFT correspondence, the membrane paradigm
correctly captures the infrared (hydrodynamic) sector of the
boundary thermal field theory. Further, we see something interesting
in its derivation. There, the diffusion constant is expressed as the
product of a factor evaluated at the horizon (\ref{one}) and an
integral from the horizon to infinity (\ref{two}). In the limit
$\lg\to{1 \ov 4}$, it is the former that approaches zero.

\end{enumerate}

\section{Causality in Bulk and on Boundary}
\label{gravitoncone}

%In this section we investigate if there might be causality
%problems in the theories we have discussed that violate the
%viscosity bound.
In this section we investigate if there are causality
problems in the bound-violating theories discussed above. First we will discuss the bulk causal structure.
Then we discuss a  curious high-momentum metastable state in  the
bulk  graviton wave equation that may have consequences for
boundary causality. The analysis in this section is refined in~\cite{newBLMSY} where we indeed see a precise signal of causality violation for $\lg>\frac{9}{100}$.

\subsection{Graviton cone tipping}
\label{conetip}

As a consequence of higher derivative terms in the gravity action,
graviton wave packets in general do not propagate on the
light cone of  a given background geometry. For example, when $\lg
\neq 0$, the equation (\ref{eom}) for the propagation of a
transverse graviton differs from that of a minimally coupled
massless scalar field propagating in the same background geometry
(\ref{bba}). To make the discussion precise, let us write (we will
consider only $x_{1, 2}$-independent waves) \be \label{envelope}
\phi(t, r, x_3)=e^{-i\omega t +i k_r r+i q x_3}\phi_{en}(t, r,
x_3). \ee Here, $\phi_{en}$ is a slowly-varying envelope function,
and we take the limit $k=(\omega, k_r, 0, 0, q)\to \infty$. In
this limit, the equation of motion (\ref{eom}) reduces to \be
\label{eikonal} k^{\mu}k^{\nu}g^{\rm eff}_{\mu\nu}\approx 0, \ \ee
where
 \be
\label{effgeo} ds_{\rm eff}^2=g^{\rm eff}_{\mu\nu}dx^{\mu}dx^{\nu}
=f(r)\ns^2 \le(-dt^2 + {1 \ov c_g^2} dx_3^2 \ri)
+\frac{1}{f(r)}dr^2.
 \ee
In (\ref{effgeo})
 \be \label{Nse}
c_g^2 (z)   = {\ns^2 \tilde f(z) \ov z^2} {1-\lg \tf'' \ov 1 -
{\lg \tf' \ov z}} \equiv c_b^2 {1-\lg \tf'' \ov 1 - {\lg \tf'
\ov z}}
 \ee
can be interpreted as the local ``speed of graviton'' on a
constant $r$-hypersurface. $c_b^2 \equiv {\ns^2 \tilde f(z) \ov
z^2} $ introduced in the second equality in~(\ref{Nse}) is the
local speed of light as defined by the background
metric~(\ref{bba}). Thus the graviton cone in general does not coincide with the
standard null cone or light cone defined by the background metric.\footnote{Note that
 \be \label{Nsee}
 {c_g^2 \ov c_b^2}   = {1-\lg \tf'' \ov 1 - {\lg \tf' \ov z}} =
{1 - 4 \lg + 12 {\lg \ov z^4} \ov 1 - 4 \lg + 4 {\lg  \ov  z^4} }
\ , \ee and in particular the ratio is greater than $1$ for $\lg >
0$. Note that bulk causality and the existence of a well-posed
Cauchy problem  do not crucially depend on reference metric
light cones and  such tipping is not a definitive sign of
causality problems. Also for any value of $\lg$, the graviton cone coincides with the
light cone in the radial direction. If not, we could have argued for
the violation of the second law of thermodynamics following
\cite{Dubovsky:2006vk,Eling:2007qd}. Further note that for $\lg <
-{1 \ov 8}$, there exists a  region outside the horizon where $c_g^2
< 0$ which will lead to the appearance of tachyonic modes, following
\cite{spectre}. We have not explored the full significance of this
instability here since it is not correlated with the viscosity bound.}  A few more comments about graviton cone are found at the end of Appendix~\ref{junk}.

\begin{figure}[t]
\includegraphics[scale=0.7,angle=0]{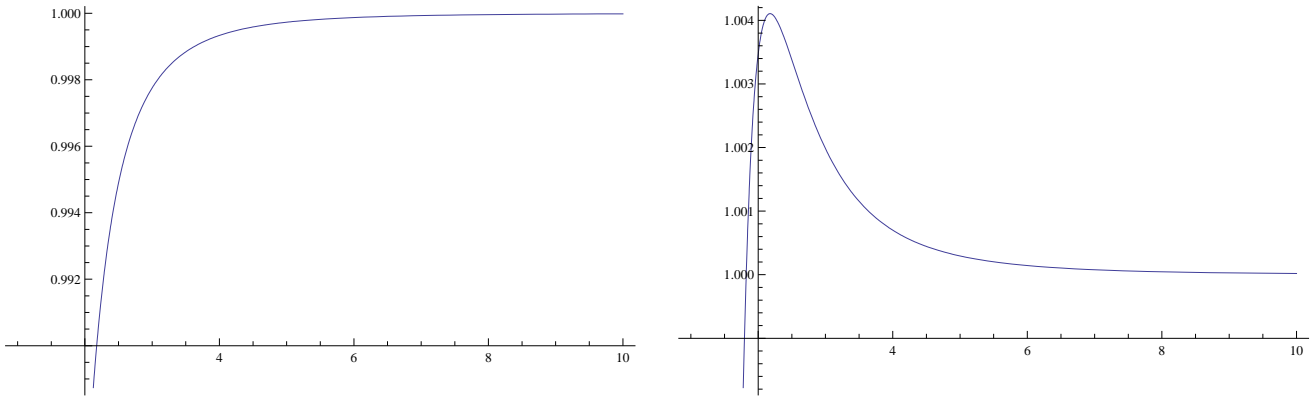}
\caption{$c_g^2 (z)$ (vertical axis) as a function of $z$
(horizontal axis) for $\lg =0.08$ (left panel) and $\lg =0.1$
(right panel). For $\lg < {9 \ov 100}$, $c_g^2$ is a monotonically
increasing function of $z$. When $\lg > {9 \ov 100}$, as one
decreases $z$ from infinity, $c_g^2$ increases from $1$ to a
maximum value at some $z>1$ and then decreases to $0$ as $z \to 1$
(horizon). }
 \label{velo}
\end{figure}

In the nongravitational boundary theory there is an invariant
notion of light cone and causality. %At a heuristic level, a
%graviton wave packet moving at its speed $c_g (z)$ in the
%bulk should translate into the boundary theory as the propagation
%of disturbances of the stress tensor at the same velocity.
At a heuristic level, a graviton wave packet moving at speed $c_g (z)$ in the
bulk should translate into disturbances of the stress tensor propagating with the same velocity in the boundary theory.
It is
thus instructive to compare $c_g$ and $c_b$ with the boundary
speed of light, which we now set to unity by taking $\ns = a$~($a$
was defined in~(\ref{Ade})). At the boundary ($z= \infty$) one finds
that $c_g (z)= c_b (z)= 1$. In the bulk, the background local
speed of light $c_b$ is always smaller than $1$, which is related
to the redshift of the black hole geometry. The local speed of graviton $c_g (z)$, however, can be greater than $1$ for
certain range of $z$ if $\lg$ is sufficiently large. To see this,
we can examine the behavior of $c_g^2$ near $z = \infty$,
 \be \label{veS}
  c_g^2 (z) - 1  =   {  b_1 \ov z^4} + O(z^{-8}) , \quad z
  \to \infty,
 \qquad b_1(\lg) = - {1 + \sqrt{1 - 4 \lg} - 20 \lg  \ov 2 (1 - 4
 \lg)} \ .
 \ee
 $b_1 (\lg)$ becomes positive and thus  $c_g^2$ increases above $1$
 if $\lg > {9 \ov 100}$. For such a $\lg$, as we decrease $z$ from
 infinity, $c_g^2$ will increase from $1$ to a maximum at some value of
 $z$ and then decrease to zero at the horizon. See Fig.~\ref{velo}
 for the plot of $c_g^2 (z)$ as a function of $z$ for two values of $\lg$.
 When $\lg = {9 \ov 100}$ one finds that
 the next order term in (\ref{veS}) is negative and thus $c_g^2$
 does not go above $1$.
 Also note that $\lg \to {1 \ov 4}$,
 $b_1 (\lg)$ goes to plus infinity.\footnote{In fact coefficients of
 all higher order terms in $1/z$ expansion become divergent in this limit.}
Thus heuristically, in the boundary theory there is a potential for superluminal propagation of disturbances of the stress tensor.

In~\cite{newBLMSY} we explore whether such bulk graviton cone
behavior can lead to boundary causality violation by studying the
behavior of graviton null geodesics in the effective geometry.
There, we indeed see causality violation for $\lg>\frac{9}{100}$.

\subsection{New metastable states at high momenta ($\lg > {9 \ov 100}$) }

We now study the behavior of the full graviton wave equation.   Let us recast the equation (\ref{eom}) in Schr\"{o}dinger form. For this purpose, we introduce
 \be
{dy \ov dz} = {1 \ov \ns \tf(z)} , \qquad \psi = B \phi , \qquad B
= \sqrt{K \ov \tf} \ .
 \ee
Then (\ref{eom}) becomes
 \be \label{enr}
 - \p_y^2 \psi + V(y) \psi = \w^2 \psi
 \ee
with
 \be \label{potential}
 V (y) = \q^2 c_g^2 (z) + V_1,  \qquad  V_1 (y) = {\p_y^2 B
\ov B} = {\ns^2 \tf^2 \ov B} \le(B'' + {\tf' \ov \tf} B' \ri) \ ,
 \ee
 where $c_g^2 (z)$ was defined in~(\ref{Nse}).
The advantage of using (\ref{enr}) is that qualitative features of
the full graviton propagation (including the radial direction) can be
inferred from the potential $V(y)$, since we have intuition for
solutions of the Schr\"{o}dinger equation. Since $y$ is
a monotonic function of $z$, below we will use the two coordinates
interchangeably in describing the qualitative behavior of $V(y)$.

\begin{figure}[t]
\includegraphics[scale=0.65,angle=0]{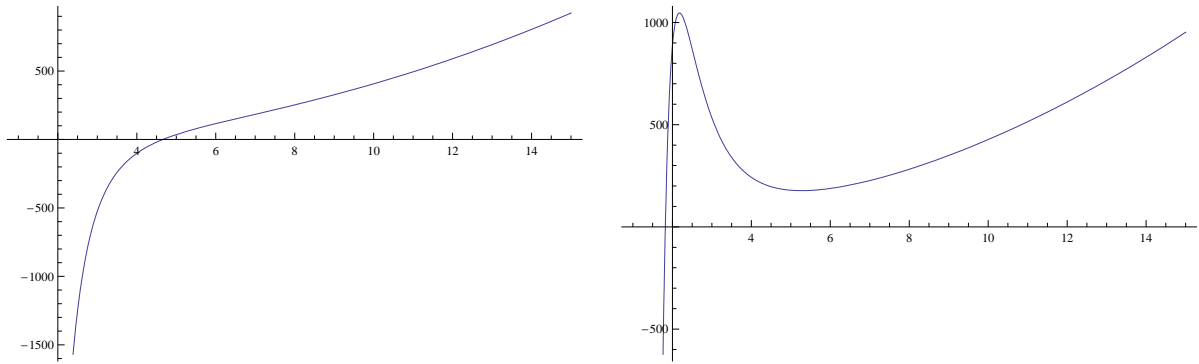}
\caption{$V(z)-q^2$ (vertical axis) as a function of $z$ (horizontal
axis) for $\lg=0.08$ and $\q =500$ (left panel) and for $\lg=0.1$
and $\q =500$ (right panel). $V(z)$ is a monotonically increasing function of
$z$ for $\lg \leq {9 \ov 100}$, but develops a local minimum for
$\lg > {9 \ov 100}$ with large enough $\q$.}
 \label{pote}
\end{figure}

One can check that $V_1 (z)$ is a monotonically increasing
function for any $\lg > 0$ (note $V_1 (z) \to + \infty$ as $z \to
\infty$). For $\lg \leq {9 \ov 100}$, $c_g^2 (z)$ is also a
monotonically increasing function as we discussed in the last
subsection and the whole $V(z)$ is monotonic. When $\lg > {9 \ov
100}$, there exists a range of $z$ where  $c_g^2 (z)$ decreases with
increasing $z$ for sufficiently large $z$. Thus $V(z)$ can now
have a local minimum for sufficiently large $\q$. For
illustration, see Fig.~\ref{pote}
 for the plot of $V (z)$ as a function $z$ for two values of $\lg$.

Generically, a graviton wave packet will fall into the black brane
very quickly, within the time scale of the inverse temperature ${1
\ov T}$ (since this is the only scale in the boundary theory).
%\footnote{There might be a loop hole to this argument, since we do
%have another dimensionless parameter $\lg$. But for $\lg \in ({9
%\ov 100}, {1 \ov 4})$, $\ns=a \sim O(1)$. Thus there does not
%appear to have much room for maneuver.}
%Such
%thermal fluctuations cannot be described as propagating
%quasi-particles  and do not lead to an apparent violation of
%causality.
Here, however, precisely when the local speed of graviton $c_g$ can exceed $1$ (i.e. for $\lg > {9 \ov 100}$), $V(z)$
develops a local minimum for large enough $\q$ and the
Schr\"{o}dinger equation~(\ref{enr}) can have metastable states
living around the minimum. Their lifetime is determined by the
tunneling rate through the barrier which separates the minimum
from the horizon. For very large $\q$ this barrier becomes very
high and an associated metastable state has lifetime
parametrically larger than the timescale set by the temperature.
In the boundary theory, these metastable states translate into
poles of the retarded Green function for $T_{xy}$ in the lower
half-plane. The imaginary part of such a pole is given by the
tunneling rate of the corresponding metastable state. Thus for
$\lg > {9 \ov 100}$, in boundary theory we find new
quasiparticles at high momenta with a small imaginary
part.\footnote{A similar type of long-lived quasiparticles exist
for $\NN=4$ SYM theory on $S^3$~\cite{guido}, but not on ${\bf
R^3}$.}

In~\cite{newBLMSY}, we confirm that those long-lived quasiparticles give rise to causality violation for $\lg>\frac{9}{100}$.

\section{Discussion}
\label{discussion}

In this paper we have computed $\eta/s $ for Gauss-Bonnet gravity using a variety of techniques.  We have found that the viscosity bound
is violated for $\lg >0$ and have looked for pathologies correlated to this violation.  For small positive $\lg$ we have not found any.
The violation of the bound becomes extreme as $\lg \rightarrow {1 \ov 4}$ where $\eta$ vanishes. We
have focused our attention on this region to find what unusual properties of the boundary theory could yield a violation not only of the bound but also of the qualitative intuitions suggesting a lower bound on $\eta/s$. Above we also have discussed a novel quasiparticle excitation. In~\cite{newBLMSY}, causality violation is firmly established for $\lg>\frac{9}{100}$.

It is also instructive to examine the
behavior of the zero temperature theory as $\lg \rightarrow {1 \ov 4}$. Basic parameters describing the boundary CFT are the coefficients of
the 4D Euler and Weyl densities called $a$ and $c$ respectively. These have been computed first in  \cite{Henningson:1998gx}, and for Gauss-Bonnet gravity in \cite{Nojiri:1999mh}. Their results indicate that
\begin{eqnarray}
\label{ac} c &\sim& (1 -4\lg)^{\ha}, \cr a &\sim& (3
(1-4\lg)^{\ha}-2).
\end{eqnarray}
The parameter $c$ is related to the two-point function of a
boundary stress tensor which is forced by unitarity to be
positive. (\ref{ac}) shows that $c$ vanishes at $\lg ={1 \ov 4}$
demonstrating the sickness of this point.\footnote{This can also
be seen from the derivations in Sec.\ref{shear}.} For $\lg$ a bit
less than ${1 \ov 4}$ the stress tensor couples very weakly in a system with a large number of degrees of freedom.  This is peculiar
indeed. In the bulk it seems that gravity is becoming strongly coupled there.

The coefficient $a$ vanishes at $\lg={5 \ov 36}$. The significance of this is unclear.

More generally, we believe it would be valuable to explore how
generic higher derivative corrections modify various gauge theory
observables. This is important not only for seeing how reliable
it is to use the infinite 't Hooft coupling approximation for
questions relevant to QCD, but also for achieving a more balanced
conceptual picture of the strong coupling dynamics.   Furthermore, this may
 generate new effective tools for separating the
swampland from the landscape.

As a cautionary note we should mention that pathologies in the boundary theory in regions that violate the viscosity bound
may not be visible in gravitational correlators, at least when $g_s = 0$.  As an example consider the $\al'^3 R^4$ terms discussed
in \cite{BLS}.   For positive $\al'$, the physical case, the viscosity bound is preserved.  But the bulk effective action can equally be
studied for $\al'$ negative.  Here gravitational correlators can be computed and will violate the viscosity bound.  The only indication
of trouble in the boundary theory at $g_s=0$ will come from correlators of string scale massive states, whose mass and CFT conformal weight
$\sim 1/(\al')^{\ha}$, an imaginary number!

\begin{acknowledgments}
We thank A.~Adams, N.~Arkani-Hamed, R-G.~Cai, A.~Dymarsky, Q.~Ejaz, T.~Faulkner, H.~Jockers,
P.~Kovtun, J.~Liu, D.~Mateos, H.~Meyer, K.~Rajagopal, D.~T.~Son,
A.~Starinets, L.~Susskind, B.~Zwiebach for discussions. HL also wishes to thank
J.~Liu for collaboration at the initial stages of the work.  We would also like to thank Yevgeny Katz and Pavel Petrov for sharing a draft of their work \cite{new}.

MB and HL are partly supported by the U.S. Department of Energy
(D.O.E) under cooperative research agreement \#DE-FG02-05ER41360.
HL is also supported in part by the A. P. Sloan Foundation and the
U.S. Department of Energy (DOE) OJI program. HL is also supported
in part by the Project of Knowledge Innovation Program (PKIP) of
Chinese Academy of Sciences. HL would like to thank KITPC
(Beijing) for hospitality during the last stage of this project.
Research at Perimeter Institute is supported by the Government of Canada through Industry Canada and by the Province of Ontario through the Ministry of Research \& Innovation. RCM also acknowledges
support from an NSERC Discovery grant and from the Canadian
Institute for Advanced Research. SS is supported by NSF grant
9870115 and the Stanford Institute for Theoretical Physics. SY is
supported by an Albion Walter Hewlett Stanford Graduate Fellowship
and  the Stanford Institute for Theoretical Physics.
\end{acknowledgments}

\appendix

\section{Thermodynamic properties of GB black holes}
\label{ap:ther}

\subsection{Free energy}
\label{freeenergy}

It is easy to confirm that the following metric is a stationary point of the Gauss-Bonnet action (\ref{action})
\begin{eqnarray}
\label{bbap} ds^2&=&-f(r)\ns^2dt^2
+\frac{1}{f(r)}dr^2+\frac{r^2}{\lad^2}
 \le(\mathop\sum_{i=1}^{3}dx_i^2 \ri) \ \ {\rm with}\cr
 f(r)&=&\frac{r^2}{\lad^2}\frac{1}{2\lg} \le(1-\sqrt{1-4\lg+4\lg \le(\frac{r_{+}}{r}\ri)^{4}}\ri) \ .
\end{eqnarray}
First note that the Hawking temperature is
\begin{equation}
\label{temperatureap}
T(r_{+})=\frac{1}{2\pi}[\frac{1}{\sqrt{g_{rr}}}\frac{d}{dr}\sqrt{g_{tt}}]|_{r=r_{+}}
=\ns\frac{1}{\pi}\frac{r_{+}}{\lad^2}.
\end{equation}

To get the free energy $F[T]$ of the
macroscopic configuration (\ref{bbap}), we note the following
correspondence in the classical limit:
\begin{equation}
e^{-\frac{1}{T}F[T]}=Z[T]=e^{-I[T]}.
\end{equation}
Here, $I[T]$ is the Euclidean action of the configuration with
temperature $T$. Evaluating the Euclideanized bulk action for
Gauss-Bonnet gravity \reef{action} with the background metric
\reef{bbap}, we find
\begin{eqnarray}
I_{\rm bulk}[T(r_{+})]&=&-\frac{1}{16\pi
G_{N}}\mathop\int_{r_{+}}^{r_{\rm max}}dr\mathop\int_{0}^{\frac{1}{T}}dt_{E}\mathop\int
d^3x_i\sqrt{g_E}[R-2\Lambda+\frac{\lambda_{GB}}{2}L^2(R^2-4R_{\mu\nu}R^{\mu\nu}+
R_{\mu\nu\rho\sigma}R^{\mu\nu\rho\sigma})] \cr &=&\frac{1}{16\pi
G_{N}}V_3\frac{\ns}{T}\frac{r_{+}^{4}}{\lad^{5}}\frac{1}{\lg} \le[\frac{r_{\rm max}^4}{r_{+}^4}\le(12\lg-5+5\sqrt{1-4\lg}\ri)-4\lg+\frac{2\lg}{\sqrt{1-4\lg}}\ri].
\end{eqnarray}
We regulate this result by subtracting the Euclidean action of the
$\lg$-modified pure AdS space (obtained by setting $r_{+}=0$ in (\ref{bbap}))
\begin{equation}
I_{\rm bulk}^{\rm pure}[T'(T(r_{+}))]=\frac{1}{16\pi
G_{N}}V_3\frac{\ns}{T'}\frac{r_{+}^{4}}{\lad^{5}}\frac{1}{\lg}\times
\le[\frac{r_{\rm max}^4}{r_{+}^4}\le(12\lg-5+5\sqrt{1-4\lg}\ri)\ri]
\end{equation}
with $T'(T)$ chosen so that the geometries at $r=r_{\rm max}$ agree
\cite{WittenThermal}. Quantitatively,
\begin{equation}
\label{matchap}
\frac{1}{T'}\sqrt{\frac{r_{\rm max}^2}{\lad^2}\frac{1}{2\lg}
\le(1-\sqrt{1-4\lg}\ri)}
=\frac{1}{T}\sqrt{\frac{r_{\rm max}^2}{\lad^2}\frac{1}{2\lg}
\le(1-\sqrt{1-4\lg+4\lg \frac{r_{+}^4}{r_{\rm max}^4}}\ri)}.
\end{equation}
All in all, we get
\begin{equation}
F[T]=T(I_{\rm bulk}[T]-I_{\rm bulk}^{\rm pure}[T'(T)])=-\frac{1}{4G_{N}}V_3(\pi
\lad T)^3 \le(\frac{T}{4}\ri)\frac{1}{\ns^3}.
\end{equation}
The entropy density is then given by
\begin{equation}
s[T]=\frac{1}{V_3}\le(-\frac{d}{dT}F[T]\ri)=\frac{1}{4G_{N}}(\pi
\lad T)^3\frac{1}{\ns^3}
=\frac{1}{4G_{N}}\le(\frac{r_{+}}{\lad}\ri)^3.
\end{equation}

\subsection{Vanishing of Gibbons-Hawking contribution}
\label{vanishingsurface} To be complete, we need to show that
there is no contribution to the free energy from the Gibbons-Hawking
surface term when we regulate with the background subtraction method presented above. This can be shown
explicitly. For the black brane solution, the Gibbons-Hawking
contribution is\footnote{A quick way to get the first equality is
to consider the action of the most general static planar symmetric
metrics, vary it, and focus on the terms involving second
derivatives. Note that with this approach, we have also accounted here for the
possible contribution of the higher derivative terms in the
generalized Gibbons-Hawking term \cite{Myers}.}:
\begin{eqnarray}
I_{GH}[T(r_{+})]&=&-\frac{1}{16\pi
G_{N}}V_3\frac{\ns}{T}\le(\frac{r}{\lad}\ri)^3 \le[6(\partial_r
f)\le(\frac{f}{r}\ri)-6\lg \le\{3\lad^2\le(\frac{\partial_r
f}{r}\ri)+2\lad^2\le(\frac{f}{r^2}\ri)\ri\}\le(\frac{f}{r}\ri)\ri]\Big|_{r=r_{\rm max}} \cr
&=&-\frac{1}{4\pi
G_{N}}V_3\frac{\ns}{T}\frac{r_{+}^4}{\lad^5}
\le(-2+3\sqrt{1-4\lg}\ri)
\le[\frac{r_{\rm max}^4}{r_{+}^4}\le(\frac{1-\sqrt{1-4\lg}}{\lg}\ri)
-\frac{1}{\sqrt{1-4\lg}}\ri] \ .
\end{eqnarray}
A similar expression is obtained for pure AdS space. With the
choice (\ref{matchap}), we obtain
\begin{equation}
I_{GH}[T]-I_{GH}^{\rm pure}[T'(T)]=0 \ .
\end{equation}

\section{$\eta/s$ from shear and sound channel poles}
\label{ap:so}

Our calculation in this appendix follows the techniques developed
in~\cite{KovtunEV}.

Consider a perturbation of the background metric of the form
$h_{\mu\nu} =h_{\mu \nu}(r) e^{- i \omega t + i q x_3}$, with
$\mu,\nu=t,r,x_1,x_2,x_3$. We can label various kinds of
perturbations according to their transformations under the
symmetry group of rotations in the $1-2$ plane. There are three
types of decoupled excitations corresponding to spin $2$ (scalar
channel), spin $1$ (shear channel) and spin $0$ (sound channel).

\subsection{Shear channel}

The shear channel excitations involve $h_{t \al}$, $h_{r \al}$ and
$h_{3 \al}$ with $\al =1,2$. Choosing the radial gauge $h_{\mu
r}=0$, the shear channel equations can be reduced to a single
equation for $Z(r)= q g^{11} h_{t1}+ \omega g^{11} h_{31}$. At
first order in $\lg$, $Z(r)$ satisfies the equation
(below we use the notations introduced in the main text,
see~(\ref{sDef}))
\begin{eqnarray}
\label{shearrsmallc} 0&=&Z''(z) + {Z'(z)\ov z} \left({5
z^4-1\ov z^4-1} +{4 \tilde q^2 \ov \tilde q^2 (-z^4+1)+z^4
{\tilde \omega^2 \ov {\ns}^2}} \right)+Z(z) \left( { \tilde
q^2 (-z^4+1)+z^4 {\tilde \omega^2 \ov {\ns}^2}\ov( z^4-1)^2
}\right)+\cr&+&{\lg \ov 2} \left[Z'(z) \left( -{8
(2\tilde q^4 (z^4-1)^2+4  \tilde q^2 z^4  {\tilde \omega^2 \ov
{\ns}^2}-3 z^8  {\tilde \omega^4 \ov {\ns}^4})\ov z^5 (
\tilde q^2 (z^4-1)-z^4 {\tilde \omega^2 \ov {\ns}^2})^2}
\right)+ Z(z) \left( 2  { \tilde q^2 (z^4 + 3 ) - 2 z^4 {\tilde
\omega^2 \ov {\ns}^2}\ov  z^4 (z^4 -1
  )}\right)\right] \ .
  \end{eqnarray}
Following a similar analysis to that at the beginning of
Sec.\ref{solution}, we find that the solution
to~(\ref{shearrsmallc}) which satisfies an infalling boundary
condition at the horizon $z=1$ can be written as
\begin{eqnarray}
\label{ans} Z(z)= \left(1-{1\ov z^4}\right)^{- i {\w \ov 4
\ns}} g(z),
\end{eqnarray}
where $g$ is regular at $z=1$. In order to find the
hydrodynamical poles,
  it is enough to  find $g (z)$  for small values
  of $\tilde \omega$ and $\tilde q$, which we will assume are of the same order.
  For this purpose, we introduce a scaled quantity
$W= {\tilde \omega \ov \tilde q {\ns}}$ and expand $g(z)$ as
a power series of $\q$. The solution can be readily found to be
\begin{eqnarray}
\label{fsol} g(z)= 1 + {i \tilde q \ov 4 W} \left(1- {1\ov
z^4} \right) \left[1+\lg \left( 3(W^2 -1) -{1\ov z^4}
\right)\right]+ O(\tilde q^2, \lg^2) \ .
\end{eqnarray}
We thus find near infinity $Z(z)$ can be expanded in $1/z$ as
 \begin{eqnarray}
\label{asym} Z(z)\approx {\cal A} + {\cal B} z^{-4} +O(z^{-8}), \quad z\rightarrow \infty,
\end{eqnarray}
where
\begin{eqnarray}
\label{abvalu} \cal{A} &=& 1+ {i\tilde q\ov 4 W} + 3 i {\tilde q
\ov 4 W} \lg \left(  W^2-1\right) +O(\tilde q^2)
\cr
 & = & 1 + {i \ns^2 \ov 4 \pi T} \le(1- 3 \lg \ri) {q^2 \ov
 \om} + {3 i \lg \om \ov 4 \pi T} + \cdots , \\
\cal{B}&=&- {i \tilde q \ov 4W}  + i {W \tilde q\ov 4}+ i
{\lg \tilde q\ov W} \left( {1\ov 2}- 3{ W^2 \ov
4}\right) +O(\tilde q^2) \cr
 & = & {i \ov 4 \pi T}{1-3 \lg \ov \om} \le(\om^2 - {\ns^2 \ov 1- \lg}
 q^2\ri) + \cdots .
\end{eqnarray}
Carrying out the procedure (\ref{bdw})--(\ref{eo}) one finds that
 \be \label{rir}
 G_R (k) \propto {\BB \ov \AA} \ .
 \ee
In particular one can show that the poles of $G_R (k)$ solely
arise from zeros of $\AA$.

The Dirichlet boundary condition corresponding to ${\cal A}=0$
determines the hydrodynamical pole as\footnote{We now need to assume
$\om \sim O(q^2)$.}
\begin{eqnarray}
\label{shearom} \omega &=& -i D q^2 +O(q^3), \qquad D = {
\ns^2 \ov 4 \pi T} \le(1- 3 \lg \ri).
\end{eqnarray}
Note that in the relation~(\ref{shearA}) between the diffusion
constant $D$ and $\eta/s$, the boundary speed of light $c$
has been set to unity (otherwise the right-hand side should be
multiplied by $c^2$). Choosing $\ns^2 = a^2 \approx 1-\lg$ (see
equation (\ref{Ade})) so that the boundary speed of light is
unity, we find that
 \begin{eqnarray}
 \label{etaovsshear}
 {\eta\ov s} =
  {1\ov 4\pi}\left( 1- 4 \lg \right)+O(\lg^2)\ .
 \end{eqnarray}

\subsection{Sound channel}

The sound channel excitations involve $h_{tt}, h_{t3}, h_{33},
h_{11}+h_{22}, h_{rr}, h_{tr}, h_{r3}$. Choosing the radial gauge
$h_{\mu r}=0$, the sound channel equations can be reduced to a
single equation for the variable
 \be
 Z_s(r)= {4 q \ov \omega} g^{33} h_{t3} +
2 g^{33} h_{33}- (g^{22} h_{22}+g^{11} h_{11})\left( 1- {q^2\ov
\omega^2} {\partial_r g_{tt} \ov \partial_r  g_{11}}\right) + 2
{q^2\ov \omega^2} {h_{tt}\ov g_{11}} \ .
 \ee
At first order in $\lg$, the equation for $Z_s(z)$ can be
written as (we use the same notation as in the main text)
\begin{eqnarray}
 &0&=Z_s''(z)+ Z_s'(z)\left({3 {\tilde
\omega^2\ov {\ns}^2} z^4 (1 - 5 z^4)+{\tilde q}^2 (9 - 16 z^4 +
15 z^8)\ov z (-1 + z^4) (-3 {\tilde \omega^2\ov {\ns}^2} z^4
+  {\tilde q}^2 (-1 + 3 z^4))} \right)+\cr
&+&Z_s(z)\left(-3 {\tilde \omega^4 \ov {\ns}^4} z^{10} + 2
{\tilde q}^2 {\tilde \omega^2\ov {\ns}^2} z^6 (-2 + 3 z^4) -
{\tilde q}^2 (-1 + z^4) (-16 + {\tilde q}^2 z^2 (-1 + 3 z^4))\ov
z^2 (-1 + z^4)^2 (-3 {\tilde \omega^2\ov {\ns}^2} z^4 +
{\tilde q}^2 (-1 + 3 z^4))\right)+\cr
&+&\lg \left[Z_s'(z)\left(4 (27 {\tilde \omega^4\ov
{\ns}^4} z^8 + 6 {\tilde q}^2 \tilde \omega^2 z^4 (-11 + z^4) +
{\tilde q}^4 (-11 + 66 z^4 - 27 z^8))\ov z^5 (-3 {\tilde
\omega^2\ov {\ns}^2} z^4 +  {\tilde q}^2 (-1 + 3
z^4))^2\right)\right.+\cr
&+&\left. {Z_s(z)\ov  z^6 (-1 + z^4) (-3
{ \tilde \omega^2\ov {\ns}^2} z^4 +  {\tilde q}^2 (-1 + 3
z^4))^2}\right. \left.\left( -18 {\tilde \omega^6\ov {\ns}^6}
z^{14}+ 3 {\tilde q}^2 {\tilde \omega^4\ov {\ns}^4} z^{10} (17
+ 15 z^4) +\right.\right.\cr
&+&\left.\left. {\tilde q}^4 ({\tilde q}^2 (7 + z^4) (z - 3 z^5)^2 +
32 (4 - 23 z^4 + 15 z^8)) - 4  {\tilde q^2 \tilde \omega^2\ov
{\ns}^2} z^4 (-180 + 132 z^4 + {\tilde q}^2 z^2 (-10 + 9 z^4 (3 +
z^4)))\right)\right].
 \nonumber
  \\ \label{soundsmallc}
\end{eqnarray}

Again the solution satisfying the infalling boundary condition at
the horizon $z=1$ can be written as
\begin{eqnarray}
\label{ans2} Z(z)= \left(1-{1\ov z^4}\right)^{- i {\w \ov 4
\ns}} s(z).
\end{eqnarray}
Defining as above the quantity $W= {\tilde \omega \ov \tilde q
{\ns}}$,  and expanding $s(z)$ in $\q$, we find that
 \begin{eqnarray}
\label{sounssol} s(z)&=& { 3 W^2 z^4- ( 1+z^4) \ov ( 3 W^2-2
)z^4} - \lg { -3 + 2 z^4 + z^8\ov z^8( 3 W^2 - 2)}
+\cr&+ &i \tilde q \left[ {W ( z^4-1) \ov z^4 ( 3 W^2-2)}+
\lg W \left(  1-{1\ov z^4}\right) {( -7 + 3 ( 3W^2
-5)z^4)\ov 4 z^4 ( 3 W^2- 2)}\right] + O(\q^2) \ .
\end{eqnarray}
The leading asymptotic behavior close to the boundary at infinity
is
$$
Z_s(z)= {\cal A}_s + {\cal B}_s z^{-4} + O(z^{-8}),
$$
with
\begin{eqnarray}
\label{asympsound} {\cal A}_s \propto q^2 (1 + \lg) - {i \ov \pi
T} q^2 \om \le(1-{15 \ov 4} \lg \ri) -{3 \om^2 \ov \ns^2}
 - {i 9 \lg\ov 4 \pi T} {\om^3 \ov \ns^2} + \cdots
\end{eqnarray}

Again, the hydrodynamical pole is found by setting ${\cal A}_s=0$, leading to
\begin{eqnarray}
\omega_{\rm sound} &=& \pm c_s q - i \Gamma_s q^2,\\
c_s&=&  {1\ov \sqrt{3}} {\ns} (1+{\lg \ov 2}),
 %= {1\ov \sqrt{3}}+{\cal O}(\lg^2)\\
 \\
\Gamma_s&=&{ 2 \ov 3}  {{\ns^2}  \ov 4 \pi T}(1- 3
\lg).
%= {2\ov 3}  {1\ov 4 \pi T}( 1-4 \lg
%)+O(\lg^2).
\end{eqnarray}
By choosing the boundary speed of light to be unity, i.e. $\ns = a
\approx (1-{\lg \ov 2}) $, we thus find that $c_s = {1 \ov
\sqrt{3}}$ and from (\ref{sound})
\begin{eqnarray}
{\eta\ov s}  = {1\ov 4\pi} (1- 4 \lg)+
O(\lg^2)\ 
\end{eqnarray}
in agreement with the results obtained from the shear channel and
the main text.

\section{Derivation of (\ref{lowEatCFT}) } \label{ap:solo}

In this appendix we give some details for obtaining $g_1 (z)$ in
equation~(\ref{lowEatCFT}). Plugging (\ref{anse}) into the equation
of motion~(\ref{eom}) one finds a  fairly complicated ordinary differential equation (ODE) for
$g_1(z)$. But, by changing variable a few times, it reduces to a
simpler one. Namely, defining
\begin{equation}
u=\sqrt{1-4\lg+4\lg \frac{1}{z^{4}}},\ \ v=1-u,
\end{equation}
we get
\begin{equation}
(1-v)(\partial_v(v\partial_v g_1 +1)) +2 (v\partial_v g_1 +1) = 0
\ .
\end{equation}
Here, we note that $-{\rm ln}(v)$ is a (singular) solution, as one
can also show from more abstract reasoning. In fact, this led to
our choice of change of variable. Now, we will solve this
equation. Defining
\begin{equation}
h_1(u)=(u-1)\partial_ug_1+1,
\end{equation}
we have
\begin{equation}
u\partial_uh_1=2 h_1,
\end{equation}
which leads to
 \be
 h_1 =c_1 u^2,
 \ee
where $c_1$ is an integration constant. Thus we find that
\begin{equation}
\label{simplestODE}
\partial_u g_1={c_1 u^2 -1 \ov u-1} = u+1 \quad {\rm choosing}
\;\; c_1 =1 \ .
\end{equation}
Note in order for $g_1 (u)$ to be nonsingular at the horizon
$u=1$, we need to choose $c_1 =1$ as we have done above. Thus we
have
 \be
  g_1 = \ha u^2 + u + c_2 \ .
  \ee
We will choose the integration constant $c_2$ so that $g_1 \to 0$
as $z \to \infty$. This then leads to~(\ref{lowEatCFT}).

\section{Stretched horizon approach}  \label{junk}

In this section, we calculate $\eta/s$ for Gauss-Bonnet gravity by
extending the stretched horizon approach of \cite{Kovtun:2003wp}
(see also \cite{Son:2007vk}). Along the way, we also explicitly
show that $\eta/s$ is independent of $\al_1$ and $\al_2$ at linear
order, as expected from the field redefinition argument in
Sec.\ref{ap:fR}. As a spin-off of this work, the framework
constructed here allows us to consider tipping of the
graviton cone in a more abstract way than that presented in
Sec.\ref{conetip}.

\subsection{Kaluza-Klein reduction}

The stretched horizon calculation of \cite{Son:2007vk} begins with
an effective Kaluza-Klein reduction of the AdS black hole metric and
treating a certain class of off-diagonal metric perturbations as a
vector in the reduced geometry. In order to develop the effective
Maxwell action for $h_{\mu y}$, we reduce along the $y$-direction:
 \be
 ds^2=\tg_{\mu\nu} dx^\mu dx^\nu + e^{2\rho}(dy+A_\mu dx^\mu)^2\ .
 \label{redux}
 \ee
To construct the theory for a higher curvature action, we need to
evaluate the various components of the Riemann tensor. This is most
efficiently done using an orthonormal frame, i.e. $ds^2=\eta_{AB}E^AE^B$, which we can conveniently choose as
\begin{eqnarray}
 E^a &=& e^a{}_\mu dx^\mu \qquad{\rm with}\  a=\htt,\hx,\hz,\hr
\nonumber\\
 E^\hy &=& e^\rho(dy + A_\mu dx^\mu) \label{frame},
\end{eqnarray}
where $e^a{}_\mu$ are some choice of tetrad components for the
reduced metric $\tg_{\mu\nu}$, which need not be specified.

Straightforward calculations then yield the following results:
\begin{eqnarray}
 R_{abcd} &=& \tR_{abcd} -{1\ov 2} e^{2\rho}\,\left(F_{a[c}F_{b|d]}
 -F_{ab}F_{cd}\right)
\nonumber\\
&=& \Rb_{abcd} -{1\ov 2} e^{2\rho}\,\left(F_{a[c}F_{b|d]}
 -F_{ab}F_{cd}\right)\ ,
\nonumber\\
 R_{a\hy b \hy} &=& -\tnab_a\tnab_b\rho-\tnab_a\rho\tnab_b\rho
 +{1\ov 4} e^{2\rho}\,F_{ac}F_{b}{}^c
\nonumber\\
&=& \Rb_{a\hy b \hy}
 +{1\ov 4} e^{2\rho}\,F_{ac}F_{b}{}^c \ ,
\nonumber\\
R_{abc \hy}&=& -{1\ov 2} e^{\rho}\,\left(\tnab_c
F_{ab}+2\,\tnab_c\rho\, F_{ab}+\tnab_b\rho\, F_{ac}-\tnab_a\rho\,
F_{bc}\right) \ .
 \label{Rcomps}
\end{eqnarray}
Our notation here is such that $\tR_{abcd}$ and $\tnab_a$ denote the
curvature components and covariant derivative of the
four-dimensional geometry specified by $\tg_{\mu\nu}$. We have also
presented the first two curvature components using the notation
$\Rb_{abcd}$ which denotes to the background curvature, i.e. the
curvature of the full five-dimensional geometry with $A_\mu=0$.
Hence, for example, $\Rb_{abc \hy}=0$.

For later convenience, we also present the components of the Ricci
tensor and scalar here:
\begin{eqnarray}
 R_{ab} &=&R^c{}_{acb}+R^\hy{}_{a\hy b}
 \nonumber\\
 &=&\tR_{ab}-\tnab_a\tnab_b\rho-\tnab_a\rho\tnab_b\rho
 -{1\ov 2}e^{2\rho}\,F_{ac}F_{b}{}^c
 \nonumber\\
 &=&\Rb_{ab}-{1\ov 2}e^{2\rho}\,F_{ac}F_{b}{}^c\,,
 \nonumber\\
R_{\hy\hy} &=&R^a{}_{\hy a\hy}=-\tnab^2\rho-(\tnab\rho)^2
 +{1\ov 4}e^{2\rho}\,F^2
 \nonumber\\
 &=&\Rb_{\hy\hy}+{1\ov 4}e^{2\rho}\,F^2\,,
 \nonumber\\
R_{a\hy} &=&R^b{}_{ab\hy}=-{1\ov 2} e^{\rho}\,\left(\tnab^a
F_{ab}+3\,\tnab^a\!\rho\, F_{ab}\right)\,,
\nonumber\\
R&=&R^a{}_a+R^\hy{}_\hy\nonumber\\
&=&\tR-2\,\tnab^2\rho-2\,(\tnab\rho)^2-{1\ov 4}e^{2\rho}\,F^2
\nonumber\\
&=&\Rb-{1\ov 4}e^{2\rho}\,F^2\,.
 \label{Rcomps2}
\end{eqnarray}

\subsection{Curvature-squared theories}

Given the above results, we can begin to apply the stretched horizon
approach to the various curvature-squared theories considered above.
First, we will confirm that for the $R^2$ and $R_{\mu\nu}R^{\mu\nu}$
theories  $\eta/s$ remains unchanged to leading order. Hence we
begin with the action \reef{epr} with $\al_3=0$:
 \be \label{epr2}
I= {1 \ov 16 \pi G_N} \int d^5 x \, \sqrt{- g} \le({12 \ov \lad^2} +
R + \lad^2  \le(\al_1 R^2+ \al_2 R_{\mu \nu} R^{\mu \nu} \ri)\ri) \
.
 \ee
The background geometry is a planar AdS black hole with metric as in
\reef{bba}:
 \be \label{background}
ds^2=-f(r)\ns^2 dt^2 +{dr^2 \ov f(r)} + {r^2 \ov \lad^2}\le( dx^2 +
dy^2 +dz^2\ri) \ ,
 \ee
where the event horizon appears at $f(r=r_+)=0$. (Note that for the
present purposes, we do not have to specify $f(r)$ in further
detail.) We introduce a metric perturbation $h^y{}_\mu=A_\mu$ and
perform a Kaluza-Klein (KK) reduction on $y$ as above. Then we wish to expand the
action \reef{epr2} to second order in the perturbation. Keeping only
the quadratic terms, the resulting action is
 \begin{eqnarray}
\label{house} I_{\rm vec}\simeq \int d^4 x \, \sqrt{- \tg}\, e^{3\rho}
\le(-{1 \ov 4} F^2 - {\lad^2\ov2}  \le[\al_1\, \Rb \,F^2 + \al_2
\le( 2\,\Rb^{ab} F_{ac}F_b{}^c -\,\Rb^{\hy \hy} F^2 -
\le(e^{-3\rho}\, \tnab^a\le(e^{3\rho}F_{ab}\ri)\ri)^2\ri)\ri]\ri)\ .
 \end{eqnarray}

Now we begin by noting that we are working perturbatively to linear
order in $\alpha_{1,2}$ and that the leading order equation of
motion for the vector perturbation is:
$\tnab^a\le(e^{3\rho}F_{ab}\ri) = O(\alpha_i)$. As a result, we
easily see that the contribution of the last term in the above
action to the equations of motion will necessarily be
$O(\alpha_i^2)$. Hence this term can be dropped in the present
analysis. Further, since the background metric \reef{background}
will satisfy Einstein's equations to leading order, $\Rb_{\mu\nu}=
-\le(4/L^2\ri)\,\tg_{\mu\nu} + O(\alpha_i)$. We can make this
replacement for the background curvatures appearing in the
$O(\alpha_i)$ terms in the action, with the result:
 \be
I_{\rm vec}\simeq \int d^4 x \, \sqrt{- \tg}\, e^{3\rho} \le(-{1 \ov
4}\, F^2 \le[1+40\al_1 - 8 \al_2  \ri]\ri) + O(\alpha_i^2)\ .
 \ee
Thus, to linear order, the only effect of these two
curvature-squared terms is to change the normalization of the
effective Maxwell action. The subsequent analysis will be
identical to that presented in \cite{Kovtun:2003wp} with the
standard result that $\eta/s = 1/4\pi$.

Next we need to construct the effective action for vector
perturbation in Gauss-Bonnet theory \reef{action}. For this purpose,
we can use the contributions calculated in the above action
\reef{house} with $\alpha_1=\lg/2$ and $\alpha_2=-2\lg$. Next we
must determine the contribution coming from the Riemann-squared
term. Using the results in \reef{Rcomps}, we have
 \begin{eqnarray}
\label{house2} I'_{\rm vec}&\simeq& \int d^4 x \, \sqrt{- \tg}\,
e^{3\rho} \lad^2\,\alpha_3 \le[ -{3\ov2}\Rb^{abcd}\,F_{ab}F_{cd} +
2\Rb^{\hy a\hy b}\,F_{ac}F_{b}{}^c\ri.
 \nonumber\\
&&\qquad\qquad\le.+\left(\tnab_c F_{ab}+2\,\tnab_c\rho\,
F_{ab}+\tnab_b\rho\, F_{ac}-\tnab_a\rho\, F_{bc}\right)^2 \ri] \ ,
 \end{eqnarray}
where in the end we will substitute $\alpha_3=\lg/2$. The first term
has already been simplified using the cyclic identity,
$R_{[abc]d}=0$. Now the second line above can be simplified by
judiciously integrating by parts, applying various identities and
using the results in \reef{Rcomps} and \reef{Rcomps2}. For example,
up to total derivatives, we have
 \begin{eqnarray}
e^{3\rho}\left(\tnab_c F_{ab}\right)^2&=& 2\,e^{-3\rho}\le(
\tnab^a\le(e^{3\rho}F_{ab}\ri)\ri)^2
+e^{3\rho}\le(\Rb^{abcd}F_{ab}F_{cd}-2\,\Rb^{ab}F_{ac}F_b{}^c\ri)
 \nonumber\\
&&\qquad+e^{3\rho}\le(4\,\tnab^a\tnab^b\rho-2\,\tnab^a\rho\,\tnab^b\rho
\ri) F_{ac}F_b{}^c \ .
\label{indent}
 \end{eqnarray}
In any event, the final result is
 \begin{eqnarray}
\label{house3} I'_{\rm vec}&\simeq& \int d^4 x \, \sqrt{- \tg}\,
e^{3\rho} \lad^2\,\alpha_3 \le[ -{1\ov2}\Rb^{abcd}\,F_{ab}F_{cd} -
2\le(\Rb^{ab}+\Rb^{\hy a\hy b}\ri)F_{ac}F_{b}{}^c\ri.
 \nonumber\\
&&\qquad\qquad\qquad\qquad\le.+3\,\Rb^{\hy\hy}\,F^2 + 2
\le(e^{-3\rho}\, \tnab^a\le(e^{3\rho}F_{ab}\ri)\ri)^2 \ri] \ .
 \end{eqnarray}
The quadratic action for the vector potential arising from the
Gauss-Bonnet theory \reef{action} is thus
 \begin{eqnarray}
\label{houseGB} I^{GB}_{\rm vec}&\simeq& \int d^4 x \, \sqrt{- \tg}\,
e^{3\rho} \le(-{1 \ov 4} F^2 - {\lg\ov4}\lad^2  \le[
\Rb^{abcd}\,F_{ab}F_{cd}
 \ri.\ri.\nonumber\\
 &&\qquad\qquad\qquad\le.\vphantom{\lg\ov4}\le.
+4\le(\Rb^{\hy a\hy b}-\Rb^{ab}\ri)F_{ac}F_{b}{}^c
+\le(\Rb-2\,\Rb^{\hy\hy}\ri)F^2 \ri]\ri) \ .
 \end{eqnarray}

\subsection{Shear viscosity via membrane paradigm}

Next we need to extend the analysis of \cite{Kovtun:2003wp} to
accommodate the generalized vector action \reef{houseGB} which
arises in Gauss-Bonnet gravity. In particular, we can write the
latter in the form
 \be
I_{\rm vec}\simeq \int d^4 x \, \sqrt{- \tg}\, \le(-{1 \ov 4}\,
F_{ab}\,X^{abcd}\,F_{cd} \ri) \ ,
 \label{general}
 \ee
where the background tensor $X^{abcd}$ necessarily has the following
symmetries:
 \be
X^{abcd}=X^{[ab][cd]}=X^{cdab}\ .
 \label{symm}
 \ee
Though this generalized action will not accommodate the
contributions of an arbitrary higher curvature term (e.g., compare
with \reef{house}), this special form would also apply for
generalized Lovelock theories of gravity. Further, in the present
case where we are studying a static AdS black brane background, this
tensor satisfies the two important properties: First, $X$ is
diagonal in the index pairs $[ab]$ and $[cd]$, e.g.,
$X^{\htt\hx\hz\hx}=0$. Second, all of the components of $X$ are
nonsingular at the horizon $r=r_+$ when described with frame
indices. Alternatively, if the tensor carries coordinate indices,
the latter can be phrased as saying that all of the components of
$X_{\mu\nu}{}^{\rho\sigma}$ are nonsingular at the horizon.

Given the above framework, one easily extends the analysis of
\cite{Kovtun:2003wp}. After defining a stretched horizon at
$r=r_H$ (with $r_H>r_+$ and $r_H-r_+\ll r_+$), the natural
conserved current to consider is
 \be
j^a =  {1\ov2} \le. \,n_b\,X^{abcd}\,F_{cd}\ri|_{r=r_H}\ ,
\label{current}
 \ee
where $n_a$ is an outward-pointing radial unit vector. One then
simply follows each of the steps appearing in \cite{Kovtun:2003wp}
to arrive at the following simple result for the effective
diffusion constant:
\begin{eqnarray}
D &=&\le.\sqrt{-\tg} \tg^{xx}\sqrt{-\tg^{tt}\tg^{rr}}
\sqrt{X_{xt}{}^{xt}\,X_{xr}{}^{xr}}\ri|_{r=r_+}\ \int_{r_+}^\infty
{(-)dr\ov \sqrt{-\tg}\, \tg^{tt}\tg^{rr}\,X_{tr}{}^{tr}}
 \nonumber\\
&=&\le.\sqrt{-\tg} \sqrt{-X^{xtxt}\,X^{xrxr}}\ri|_{r=r_+}\
\int_{r_+}^\infty {(-)dr\ov \sqrt{-\tg}\,X^{trtr}}\ .
 \label{diffusion}
\end{eqnarray}
For the standard effective Maxwell action with Lagrangian $-{1\ov4
g_{\rm eff}^2} F^2$,
 \be
X_{xt}{}^{xt}= {1\ov 2\,g_{\rm eff}^2}=X_{xr}{}^{xr}=X_{tr}{}^{tr}
\nonumber
 \ee
and then the first expression above reduces to the usual result,
first derived in \cite{Kovtun:2003wp}.

Now we apply this analysis to the specific action \reef{houseGB}
arising from Gauss-Bonnet gravity. First we must extract expressions
for the background tensor, which is simplified if we divide up $X$
into three contributions
 \be
X_{ab}{}^{cd}=\X 0abcd+\X 1abcd+\X 2abcd\ , \label{deaf}
 \ee
where $X^{\ssc(0)}$ and $X^{\ssc(1)}$ correspond to the
contributions coming from the terms proportional to $F^2$ and
$F_{ac}F_{b}{}^c$, respectively. $X^{\ssc(2)}$ captures the
remaining contributions. For the action \reef{houseGB}, one finds
\begin{eqnarray}
\X 0abcd&=&\delta_{[a}{}^c\,\delta_{b]}{}^d\,e^{3\rho} \le( 1 +
\lg\lad^2
\le(\Rb-2\Rb^{\hy\hy}\ri)\ri), \nonumber\\
\X 1abcd&=&Y_{[a}{}^{[c}\,\delta_{b]}{}^{d]}\,e^{3\rho}\qquad{\rm
with}\ Y_a{}^b=4\lg\lad^2\le(\Rb_{\hy a}{}^{\hy b}-\Rb_a{}^b\ri),
\nonumber\\
\X 2abcd &=&\lg\lad^2\,e^{3\rho}\,\Rb_{ab}{}^{cd}\ .
 \label{deaf2}
\end{eqnarray}

The expression \reef{diffusion} for the diffusion constant requires
three of the components of $X$ in particular. Using the background
metric \reef{background} and the expressions \reef{deaf2}, one finds
that
\begin{eqnarray}
X_{tr}{}^{tr}&=&{1\ov2} e^{3\rho}\le(1-2\lg \le({\lad^2 \ov r^2}f\ri)\ri), \nonumber\\
X_{xr}{}^{xr}=X_{xt}{}^{xt}&=&{1\ov2}e^{3\rho}\le(1-\lg {\lad^2\ov
r} \partial_r f\ri)\ .
 \label{result}
\end{eqnarray}
To proceed further, we must explicitly introduce the solution
\reef{perturb}
 \be
{\lad^2 \ov r^2}f(r) =
{1\ov2\lg}\le[1-\sqrt{1-4\lg\le(1-{r_+^{\,4}\ov r^4}\ri)}\ri]
 \label{explicit}
 \ee
for the black brane in the Gauss-Bonnet theory. Recall that the
temperature \reef{temperature} is given by $T=\ns\,r_+/\pi \lad^2$.
Further implementing the KK reduction \reef{redux} on this
background \reef{background} yields
 \be
\sqrt{-\tg}=\ns{r^2\ov \lad^2} ,\qquad e^{3\rho}={r^3\ov \lad^3}\ .
 \label{result2}
 \ee
Given these results, the prefactor in \reef{diffusion} reduces to
 \be
\le.\sqrt{-\tg} \tg^{xx}\sqrt{-\tg^{tt}\tg^{rr}}
\sqrt{X_{xt}{}^{xt}\,X_{xr}{}^{xr}}\ri|_{r=r_+} =
\le.X_{xt}{}^{xt}\ri|_{r=r_+} =
{1\ov2}\,{r_+^3\ov\lad^3}\,\le(1-4\lg\ri)\ .
 \label{one}
 \ee

We note that the second factor in $X_{tr}{}^{tr}$ has a particularly
simple form: $1-2\lg {\lad^2 \ov r^2}f =\sqrt{1-4\lg\le(1-{r_+^4\ov r^4}\ri)}$. Then
the integral in \reef{diffusion} is evaluated as
 \begin{eqnarray}
\int_{r_+}^\infty {(-)dr\ov \sqrt{-\tg}\,
\tg^{tt}\tg^{rr}\,X_{tr}{}^{tr}}&=&2\lad^5\ns\int_{r_+}^\infty
{dr/r^5\ov \sqrt{1-4\lg\le(1-{r_+^4\ov r^4}\ri)}}
 \nonumber\\
 &=& {\lad^5\ov 2r_+^4}\,\ns\,{1-\sqrt{1-4\lg}\ov2\lg}\ .
 \label{two}
 \end{eqnarray}
Combining the results in \reef{one} and \reef{two} then yields
 \be
D
={\lad^2\ov4r_+}\le(1-4\lg\ri)\,\ns\,{1-\sqrt{1-4\lg}\ov2\lg}={c^2\ov
4\pi\ T}\,\le(1-4\lg\ri)\ ,
 \label{diffusion2}
 \ee
where $c=\ns/a$ is the boundary speed of light, with $a$ defined in
\reef{Ade}. Hence we recover the expected result for the
Gauss-Bonnet theory:
 \be
{\eta\ov s}={D T\ov c^2}={1\ov 4\pi}\,\le(1-4\lg\ri)\ .
 \label{hurrah}
 \ee

\subsection{Graviton cone revisited} \label{hockey}
Given that the background tensor $X$ is expressed in terms of
curvatures of the background spacetime (see \reef{deaf} and
\reef{deaf2}), we should be able to express the effective ``null''
cone of the gravitons in terms of these curvatures.

In Sec.\ref{conetip}, we considered the scalar channel which
corresponds to a perturbation $h_x{}^y$ with dependence on $t$, $r$
and $z$. In the present notation, this is precisely an excitation
of the vector component $A_x$. In a high-frequency or WKB limit, we
write
 \be
A_x = e^{ik\cdot x}\  \phi_{en}\ ,\label{WKBee}
 \ee
where the first factor is the rapidly varying phase and $\phi_{en}$
is the slowly modulated envelope function. The coordinate dependence of the scalar
channel also requires that $k_x=0=k_y$. For these modes, the
effective action \reef{general} reduces to
 \be
I_{\rm vec}\simeq \int d^4 x \, \sqrt{- \tg}\, \le(-\, F_{a\hx}\,X^{a\hx
b\hx}\,F_{b\hx} \ri) \ .
 \label{effect}
 \ee
From this action, we can readily derive the full equations of
motion, however, we do not need these here. In the high-frequency
limit, the graviton cone is given by
 \be
0=X^{a\hx b\hx}\,k_a\,k_b\ .
 \label{cone}
 \ee
Now as indicated above, we use \reef{deaf} and \reef{deaf2} to
express this result in terms of the background curvatures. Hence the
effective metric defining the graviton cone can be written as
 \begin{eqnarray}
2 e^{-3\rho}\,X_{a\hx}{}^{b\hx}&=& \delta_a{}^b
-2\lg\lad^2\le(\Rb_a{}^b-{1\ov2}\Rb\,\delta_a{}^b\ri)
 \nonumber\\
&&\qquad\qquad +2\lg\lad^2\le(\Rb_{a\hy}{}^{b\hy} +
\Rb_{a\hx}{}^{b\hx} -2\,\Rb_{\hy\hx}{}^{\hy\hx}\,\delta_a{}^b\ri),
 \label{effmetric}
 \end{eqnarray}
where we have implicitly assumed that $a$ and $b$ only take values
in $\lbrace \htt,\hr,\hz \rbrace$. We have also canceled certain
terms using $R_\hx{}^\hx=R_\hy{}^\hy$ for the backgrounds of
interest here. In the first line, the correction term is
proportional to the Einstein tensor, a result that is reminiscent of
that in \cite{Aragone,choquet}. Their results for the characteristic
hypersurfaces of Gauss-Bonnet gravity do not include the nontrivial
contribution in the second line above. We do not entirely understand
the source of this discrepancy but note that the analysis of
\cite{Aragone,choquet} uses complementary techniques to ours. At
least in the context of small $\lg$, the additional terms in
\reef{effmetric} have an important consequence. That is, using the
equations of motion for the background geometry, the results of
\cite{Aragone} would have predicted that the deviation of the
graviton cone from the standard light cone only occurs at
$O(\lg^2)$. However, our results in Sec.\ref{gravitoncone} indicate
that there is nontrivial result at $O(\lg)$. The additional terms
appearing in the second line of \reef{effmetric} must be responsible
for this effect.

\end{document}